\documentclass[11pt]{iopart}
\usepackage[dvipdf]{graphicx}

\usepackage{amssymb,amsfonts,iopams}

\def\av#1{\langle#1\rangle}
\newcommand{\meqref}[1]{{\bf(\hskip 1pt #1\hskip 1pt)}}
\newcommand{\eqref}[1]{(\hskip 1pt \ref{#1}\hskip 1pt)}

\begin{document}







\title{Stochastic fluctuations in metabolic pathways}


\author{Erel Levine and Terence Hwa}
\address{Center for Theoretical Biological Physics and Department of
Physics, University of
California at San Diego, La Jolla, CA 92093-0374}




\begin{abstract}
  Fluctuations in the abundance of molecules in
the living cell may affect its growth and well being.
For regulatory molecules (e.g., signaling proteins or transcription
factors), fluctuations in their expression can affect the levels
of downstream targets in a network.
Here, we develop an analytic framework to investigate the phenomenon of
noise correlation in molecular networks. Specifically, we focus on
the metabolic network, which is highly inter-linked, and noise
properties may constrain its structure and function.
Motivated by the analogy between the dynamics of a linear metabolic pathway
and that of the exactly soluable linear queueing network
or, alternatively, a mass transfer system,
we derive a plethora of results concerning fluctuations in the abundance
of intermediate metabolites in various common motifs of the metabolic network.
For all but one case examined, we find the steady-state fluctuation
in different nodes of the pathways to be effectively uncorrelated.
Consequently,  fluctuations in enzyme levels only affect local properties
and do not propagate elsewhere into metabolic networks,
 and intermediate metabolites can be freely shared
by different reactions. Our approach may be
applicable to study metabolic networks with more complex topologies,
or protein signaling networks which are governed by similar
biochemical reactions.  Possible
implications for bioinformatic analysis of metabolimic data are
discussed.
\end{abstract}

\maketitle

Due to the  limited number of molecules for typical
molecular species in microbial cells, random fluctuations in
molecular networks are common place and may  play important roles in
vital cellular processes. For example, noise in sensory signals can
result in pattern formation and collective dynamics \cite{Zhou05},
and noise in signaling pathways can lead to cell-to-cell variability
\cite{Brent05}. Also, stochasticity in gene expression has
implications on cellular regulation \cite{Raser05,Collins05} and may
lead to phenotypic diversity \cite{Kussel05,Suel06}, while
fluctuations in the levels of (toxic) metabolic intermediates may
reduce metabolic efficiency \cite{Fell97} and impede cell growth.

In the past several years, a great deal of experimental and
theoretical efforts have focused on the stochastic expression of
{\em individual} genes, at both the translational and
transcriptional levels  \cite{Swain02,Pedraza05,Golding05}.   The
effect of stochasticity on networks has been studied in the context
of small, ultra-sensitivie genetic circuits, where noise at a
circuit node (i.e., a gene) was shown to either attenuate or amplify
output noise in the {\em steady state} \cite{Tattai02,Weiss05}. This
phenomenon
--- termed {\em `noise propagation'} ---   make the steady-state fluctuations at
one node of a gene network dependent in a complex manner on
fluctuations at other nodes, making it difficult for the cell to
control the noisiness of individual genes of interest
\cite{Weiss06}. Several key questions which arise from these studies
of genetic noise include (i) whether stochastic gene expression
could further propagate into signaling and metabolic networks
through fluctuations in the levels of key proteins controlling those
circuits, and (ii) whether noise propagation occurs also in those
circuits.

Recently, a number of approximate analytical methods have been
applied to analyze small genetic and signaling circuits; these
include the independent noise approximation
\cite{Paulsson04,Shibata05,Simpson06}, the linear noise
approximation \cite{Paulsson04,Wolde06}, and the self-consistent
field approximation \cite{Sasai03}. Due perhaps to the different
approximation schemes used, conflicting conclusions have been
obtained regarding the extent of noise propagation in various
networks (see, e.g., \cite{Wolde06}.) Moreover, it is difficult to
extend these studies to investigate the dependences of noise
correlations on network properties, e.g., circuit topology, nature
of feedback, catalytic properties of the nodes, and the
parameter dependences (e.g., the phase diagram). It is of course also
difficult to elucidate these dependences using numerical simulations
alone, due to the very large degrees of freedoms involved for a
network with even a modest number of nodes and links.

In this study, we describe
an analytic approach to characterize the probability distribution for {\em all} nodes
of a class of molecular networks in the steady state. Specifically, we apply the method
to analyze  fluctuations and their correlations in metabolite concentrations
for various core motifs of the metabolic network.
The metabolic network consists of nodes which are the metabolites,
linked to each other by enzymatic reactions that convert one
metabolite to another. The predominant motif in the metabolic
network is a linear array of nodes linked in a given direction (the
directed pathway), which are connected to each other via converging
pathways and diverging branch points \cite{Michal}. The activities
of the key enzymes are regulated allosterically by metabolites from
other parts of the network, while the levels of many enzymes are
controlled transcriptionally and are hence subject to deterministic
as well as stochastic variations in their expressions \cite{Berg}. To understand the control of metabolic network, it
is important to know how changes in one node of the network affect
properties elsewhere.

Applying our analysis to directed linear metabolic pathways,
we predict that  the distribution of molecule
number of the  metabolites at intermediate nodes to be {\em statistically
independent} in the steady state, i.e., the noise does not propagate.
Moreover, given the properties of the enzymes in the pathway and the input flux,
we provide a recipe which specifies the exact metabolite distribution function
at each node. We then show that the method can be extended 
to linear pathways with reversible links, with feedback control, to cyclic and
certain converging  pathways, and even
to pathways in which flux conservation is violated (e.g., when metabolites
leak out of the cell). 
We find that in these cases correlations between nodes are negligable or vanish completely,
although nontrivial fluctuation and correlation do dominate
for a special type of converging pathways.
Our results suggest that for vast parts of the metabolic network,
different pathways can be coupled to each other
without generating complex correlations, so that properties of one node
(e.g., enzyme level) can be changed over a broad range
without affecting behaviors at other nodes.
We expect that the realization of this remarkable property will
shape our understanding of the operation of the metabolic network,
its control, as well as its evolution.
For example, our results suggest that correlations between steady-state fluctuations
in different metabolites bare no information on the network structure.
In contrast, temporal propagation of the response to an external perturbation
should capture - at least locally - the morphology of the network.  Thus, the topology
of the metabolic network should be studied during transient periods of relaxation
{\em towards} a steady-state, and not {\em at} steady-state.

Our method is motivated by the analogy between the dynamics of biochemical
reactions in metabolic pathways
and that of the exactly solvable queueing
systems \cite{Taylor98} or, alternatively, as mass transfer systems
\cite{Liggett,Levine05}. Our approach may be applicable also to analyzing
fluctuations in signaling networks, due to the close analogy
between the molecular processes underlying the metabolic and signaling
networks.
To make our approach accessible to a broad class of circuit modelers
and bioengineers who may not be familiar with
nonequilibrium statistical mechanics,
we will present in the main text only the mathematical results supported by
stochastic simulations, and defer derivations and illustrative calculations
to the Supporting Materials. While our
analysis is general, all examples are taken from amino-acid
biosynthesis pathways in {\it E. Coli} \cite{Coli}.

%
%

%
\section{Individual Nodes}

\subsection{A molecular Michaelis-Menton model}

In order to set up the grounds for analyzing a reaction pathway and
to introduce our notation, we start
by analyzing fluctuations in a single metabolic reaction catalyzed by an enzyme.

Recent advances in experimental techniques have made it possible to
track the enzymatic turnover of a substrate to product at the
single-molecule level \cite{Xie99,English05}, and to study
instantaneous metabolite concentration in the living cell
\cite{Arkin97}.
To describe this fluctuation mathematically, we model
the cell as a reaction vessel of volume $V$, containing
$m$ substrate molecules ($S$) and $N_E$ enzymes ($E$). A single
molecule of $S$
 can bind to a single enzyme $E$ with rate $k_+$ per volume,
and form a complex, $SE$. This complex, in turn, can unbind (at rate
$k_-$) or convert $S$ into a product form, $P$, at
rate $k_2$. This set of reactions is summarized by
\begin{equation}
S+E \mathop{\leftrightarrows}^{k_+}_{k_-} SE \mathop{\rightarrow}^{k_
{2}} P+E\;.
\label{e.mm}
\end{equation}
Analyzing these reactions within a mass-action framework --- keeping
the substrate concentration
fixed, and assuming fast equilibration between the substrate and the
enzymes
$(k_\pm \gg k_2)$ ---  leads to the Michaelis-Menten (MM) relation between
the macroscopic flux $c$ and the substrate concentration $[S]=m/V$ :
\begin{equation}
c = v_{\max} [S] / ([S]+K_M)\;,\label{e.mmeq}
\end{equation}
where
  $K_M=k_-/k_+$ is the dissociation
constant of the substrate and the enzyme, and $v_{\max}=k_2 [E]$
is the maximal flux, with $[E]=N_E/V$ being the total enzyme concentration.

Our main interest is in noise properties, resulting from the
discreteness of molecules. We therefore need to track individual
turnover events. These are described by the  turnover rate $w_m$,
defined as
the inverse of the mean waiting time per volume between the
(uncorrelated\footnote{We note in passing that some correlations do
exist -- but not dominate -- in the presence of ``dynamical
disorder'' \cite{English05}, or if turnover is a multi-step process
\cite{Kou05,Qian02}.}) synthesis of one product molecule to the next.
Assuming again fast equilibration between the substrate and the
enzymes, the probability of having $N_{SE}$ complexes given $m$
substrate molecules and $N_E$ enzymes is simply given by the
Boltzmann distribution,
\begin{equation} p(N_{SE}|m,N_E) =
\frac{K^{-N_{SE}}}{Z_{m,N_E}} \frac{m!
N_E!}{N_{SE}!(m-N_{SE})!(N_E-N_{SE})!}\;\label{e.boltzman}
\end{equation}
for $N_{SE}<N_E$ and $m$. Here $K^{-1}=Vk_+/k_-$  is  the Boltzmann
factor 
associated with the formation of an SE complex,
and the $Z_{m,N_E}$ takes care of normalization (i.e., chosen such that
$\sum_{N_{SE}} \, p(N_{SE}|m,N_E) =1$.)
Under this condition, the
turnover rate $w_m = \frac{k_2}{V} \sum {N_{SE}}\cdot p(N_{SE}|m,N_E)$
is given approximately by
\begin{equation}
w_m = v_{\max} \frac{m}{m+(K+N_E-1)}+{\cal O}(K^{-3})\;,
\label{e.wrate}
\end{equation}
with $v_{\max}=k_2 N_E/V$; see Supp. Mat.
We note that for a single enzyme ($N_E=1$), one has
$w_m = v_{\max}{m}/{(m+K)}$,
which was derived and verified experimentally~\cite{English05, Kou05}.

\subsection{Probability distribution of a single node}
In a metabolic pathway, the number of substrate molecules is not
kept fixed; rather, these molecules are synthesized or imported
from the environment, and at the same time turned over into
products. We consider the influx of substrate molecules to be a
Poisson process with rate $c$. These molecules are turned into
product molecules with rate $w_m$ given by Eq.~\eqref{e.wrate}.
The number of substrate molecules is now fluctuating, and one can ask
what is the probability $\pi(m)$ of finding $m$ substrate molecules
at the steady-state. This probability can be found by solving the
steady-state Master equation for this process (see Supp. Mat.), yielding
\begin{equation}
\pi(m) = {m+K+(N_E-1)\choose m}(1-z)^{K+N_E}z^m\;,
\label{e.pi}
\end{equation}
where $z = {c}/v_{\max}$ \cite{Elf03}.
The form of this distribution is plotted in supporting figure~1 (solid black line).
As expected, a
steady state exists only 
when ${c} \leq v_{\max}$.
Denoting the steady-state average by angular brackets,
i.e., $\av{x_m} \equiv \sum_m  x_m\, \pi(m)$, the condition
that the incoming flux equals the outgoing flux is written as
\begin{equation}
c = \av{w_m} = v_{\max}\frac{s}{s+(K+N_E)}\;,
\label{e.currentdensity}
\end{equation}
where $s\equiv\av{m}$.

Comparing this microscopically-derived
flux-density relation with the MM relation \eqref{e.mmeq} using the
obvious correspondence $[S]=s/V$, we see that the two are equivalent
with $K_M = (K+N_E)/V$. Note that this microscopically-derived form
of MM constant is {\em different} by the amount $[E]$ from the
commonly used (but approximate form)
$K_M=K/V$, derived from mass-action. However, for
typical metabolic reactions, $K_M \sim 10-1000\,\mu M$ \cite{Coli}
while $[E]$ is not more than 1000 molecules in a bacterium cell
($\sim 1 \mu M$); so the numerical values of the two expressions
may not be very different.

We will characterize the variation of substrate concentration in the steady-state
by the noise index
\begin{equation}
\label{e.etas}
\eta_s^2 \equiv \frac{\sigma_s^2}{s^2} = \frac{v_{\max}}{c \cdot (K
+N_E)} \;, 
\end{equation}
where $\sigma_s^2$ is the variance of the distribution $\pi(m)$.
Since $c \le v_{\max}$ and increases with $s$ towards $1$
(see Eq.~\ref{e.currentdensity}),\;
$\eta_s$ decreases with the average occupancy $s$ as expected.
It is bound from below by $1/\sqrt{K+N_E}$, which can easily be several
percent. Generally, large noise is obtained when the reaction is
catalyzed by a samll number of high-affinity enzymes (i.e., for low $K$ and $N_E$).

\section{Linear pathways}
\subsection{Directed pathways}
\begin{figure}[t]
\centerline{{\includegraphics{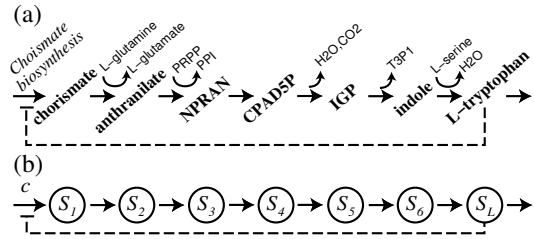}} }\caption{Linear
biosynthesis pathway. (a) Tryptophan biosynthesis pathway in {\it
E. Coli}. (b) Model for a directed pathway. Dashed lines depict
end-product inhibition. Abbreviations: CPAD5P, 1-O-Carboxyphenylamino 1-deoxyribulose-5-phosphate; NPRAN, N-5-phosphoribosyl-anthranilate; IGP, Indole glycerol phosphate;
PPI, Pyrophosphate; PRPP, 5-Phosphoribosyl-1-pyrophosphate; T3P1, Glyceraldehyde 3-phosphate.
} \label{f.pathway}
\end{figure}

We now turn to a directed
metabolic pathway, where an incoming flux of  substrate molecules is
converted, through a series of enzymatic reactions, into a product
flux \cite{Michal}. Typically, such a pathway involves the order of
10 reactions, each takes as precursor the product of the preceding
reaction, and frequently involves an additional side-reactant (such as a
water molecule or ATP) that is abundant in the cell (and whose
fluctuations can be neglected). As a concrete example, we
show in figure~\ref{f.pathway}(a) the tryptophan biosynthesis pathway
of {\it E. Coli} \cite{Coli}, where an incoming flux of chorismate
is converted through 6 directed reactions into an outgoing flux of
tryptophan, making use of several side-reactants.
Our description of a linear pathway includes an incoming flux $c$ of
substrates of type  $S_1$ along with a set of reactions that convert
substrate type $S_i$ to $S_{i+1}$ by enzyme $E_i$ (see figure~\ref{f.pathway}(b))
with rate $w^{(i)}_{m_i}=v_i m_i / (m_i+K_i-1)$ according to
Eq. \eqref{e.wrate}. We denote the number of molecules of
intermediate $S_i$ by $m_i$, with $m_1$ for the substrate and $m_L$ for the
end-product. The superscript $(i)$ indicates explicitly that the
 parameters $v_i=k_2^{(i)} N_E^{(i)}/V$ and $K_i=(K^{(i)}+N_E^{(i)})$
describing the  enzymatic reaction $S_i \to S_{i+1}$ are expected
to be different for different reactions.

The steady-state of the pathway is fully described by the joint
probability distribution $\pi(m_1,m_2,\ldots,m_L)$ of having $m_i$
molecules of intermediate substrate type $S_i$.
Surprisingly, this  steady-state distribution is given {\em exactly} by a
product measure,
\begin{equation}
\pi(m_1,m_2,\ldots,m_L)= \prod_{i=1}^{L}\pi_i(m_i)\;,
\label{e.product}
\end{equation}
where $\pi_i(m)$ is as given in Eq. \eqref{e.pi} (with $K+N_E$
replaced by $K_i$ and $z$ by $z_i = {c}/{v_i}$), as we show in Supp. Mat.
This result indicates that in
the steady state, the number of molecules of one intermediate is
statistically independent of the number of molecules of any other
substrate\footnote{We note, however, that short-time correlations
between metabolites can still exist, and may be probed for example by
measuring two-time cross-correlations; see discussion at the end of
the text.}.
The result has been derived  previously in the context of
queueing networks  \cite{Taylor98},
and of mass-transport systems \cite{Levine05}. Either may serve as a
useful analogy for a metabolic pathway.

\begin{figure*}[t]
{\includegraphics{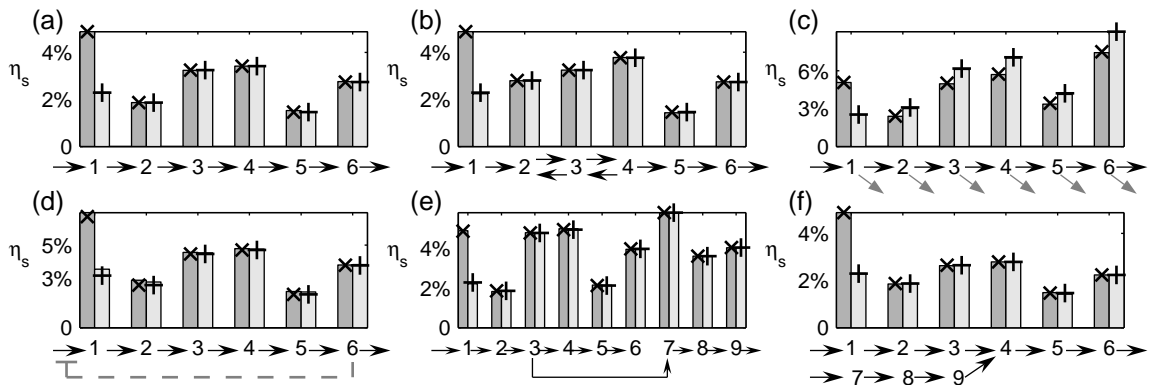}} \caption{Noise in metabolite molecular number ($\eta_s = \sigma_s/s$) for different pathways. Monte-Carlo simulations (bars) are compared with the analytic prediction (symbols) obtained by assuming decorrelation for different nodes of the pathways.  The structure of each pathway is depicted under each panel. Parameter values were chosen randomly such that $10^3<K_i<10^4$ and $c<v_i<10c$.  SImilar decorrelation was
obtained for $100$~different random choices of parameters, and for $100$~different sets with
$K_i$ 10-fold smaller (data not shown).
The effect on the different metabolites of a change in the velocity of the first reaction, $v_1=1.1c$ (dark gray)$ \to 5c$ (light gray), is demonstrated.  Similar results are obtained for changes in $K_1$ (data not shown.) (a) Directed pathway. Here the decorrelation property is exact. (b) Directed pathway with two reversible reactions. For these reactions, $v^+_{3,4}=8.4, 6.9 c;  v^-_{3,4}=1.6, 3.7,  c;  K_{3,4}^+=2500, 8000$ and $K_{3,4}^-=7700, 3700$. (c) Linear dilution of metabolites. Here $\beta/c = 1/100$. (d)  End-product inhibition,where the influx rate is given by $\alpha = {c_0}\left[{1+ (m_L/K_I)}\right]^{-1}$   with $K_I = 1000$. (e) Diverging pathways. Here metabolite $4$ is being processed by two enzymes (with different affinities, $K^\mathtt{I}=810, K^\mathtt{II}=370$) into metabolites $5$ and $7$, resp. (f) Converging pathways. Here two independent 3-reaction  pathways , with fluxes $c$ and $c'=c/2$, produce the same product, $S_4$. }
\label{f.bars}
\end{figure*}

Since the different metabolites in a pathway are statistically
{\it decoupled} in the steady state,  the mean $s_i=\av{m_i}$
and the noise
index $\eta_{s_i}^2 = c^{-1}v_i/K_i$ can be determined
by Eq. \eqref{e.etas}  individually for each node of the pathway.
It is an interesting consequence of  the decoupling
property of this model  that both the mean concentration of each
substrate and the fluctuations depend only on the properties of
the enzyme immediately downstream. While the steady-state
flux $c$ is a constant throughout the pathway, the parameters $v_i$
and $K_i$ can be set separately for each reaction
by the copy-number and  kinetic properties of
the enzymes (provided that $c<v_i$).
Hence, for example, in a case where a specific
intermediate may be toxic, tuning the enzyme properties may serve
to decrease fluctuations in its concentration, at the price of a
larger mean.  To illustrate the decorrelation between different metabolites,
we examine the response of steady-state fluctuations to a  5-fold increase in
the enzyme level $[E_1]$.
Typical time scale for changes in enzyme level much exceeds those of  the enzymatic reactions.
Hence, the enzyme level changes may be considered as quasi-steady state.
In figure~\ref{f.bars}(a) we plot the noise indices of the different metabolites.
While noise in the first node is significantly reduced upon a 5-fold increase in $[E_1]$,
fluctuations at the other nodes are not affected at all.

\subsection{Reversible reactions}
The simple form of the steady-state distribution \eqref{e.product}
for the directed pathways
may serve as a starting point to obtain additional results
for metabolic networks with more elaborate features.
We demonstrate such applications of the method by
some examples below. In many pathways,
some of the reactions are in fact reversible. Thus, a
metabolite $S_i$ may be converted to metabolite $S_{i+1}$ with rate
$v_{\rm max}^+m_i/(m_i+K_i^+)$ or to substrate $S_{i-1}$ with rate
$v_{\rm max}^-m_i/(m_i+K_i^-)$. One can show ---  in a way similar to
Ref.~\cite{Levine05} --- that the decoupling property \eqref{e.pi}
holds exactly only if the ratio of the two rates is a constant independent of $m_i$, i.e.
when $K_i^+=K_i^-$. In this case  the steady state probability is
still given by~\eqref{e.pi}, with the local currents obeying
\begin{equation}\label{e.balance}
v_i^+ z_i - v_{i+1}^-z_{i+1} = c\,.
\end{equation}
This is nothing but  the simple fact that the overall flux is the
difference between the local current in the direction of the pathway
and that in the opposite direction.

In general, of course,  $K_i^+ \neq K_i^-$.  However, we expect
the distribution to be given approximately by the product
measure  in the following situations:  (a) $K_i^+ \simeq K_i^- $;
(b) the two reactions are in the zeroth-order regime, $s\gg
K_i^\pm$ ; (c) the
two reactions are in the linear regime, $s\ll K_i^\pm$. In the
latter case Eq. \eqref{e.balance} is replaced by
$\frac{v_i^+}{K_i^+} z_i - \frac{v_{i+1}^-}{K_{i+1}^-}z_{k+1} = c\,.$
Taken together, it is only for a narrow region (i.e., $s_i \sim K_i$)
where the product measure may not be applicable.
This prediction is tested numerically, again by comparing two
pathways (now containing reversible reactions) with 5-fold difference
in the level of the first enzyme. From figure\ref{f.bars}(b), we see again
that the difference in noise indices exist only in the first node, and the
computed value of the noise index at each node is in excellent agreement
with predictions based on the product measure (symbols). SImilar decorrelation was
obtained for $100$~different random choices of parameters, and for $100$~different sets with
$K_i$ 10-fold smaller (data not shown).

\subsection{Dilution of intermediates} In the
description so far, we have ignored possible catabolism of
intermediates or dilution due to growth. This makes the flux a
conserved quantity throughout the pathway, and is the basis of the
flux-balance analysis \cite{Palsson02}. One can generalize our
framework for the case where flux is not conserved, by allowing
particles to be degraded with rate $u_m$. Suppose, for example, that
on top of the enzymatic reaction a substrate is subjected to an
  effective linear degradation,  $u_m = \beta m$. This includes the
effect of
dilution due to growth, in which case $\beta = \ln(2)/$(mean cell
division time), and the effect of leakage out of the cell.
As before, we first consider the dynamics at a single node, where
the  metabolite is randomly produced (or transported) at a rate
$c_0$. It is straightforward to generalize the Master equation for the microscopic process
to include $u_m$, and solve it in the same way.
With $w_m$ as before, the steady state distribution of the substrate
pool size is then found to be
\begin{equation}\label{e.pib}
\pi(m) =  \frac1Z{m+K -1\choose
m}\frac{(c_0/\beta)^m}{(v/\beta+K)_m}\;,
\end{equation}
where $(a)_m \equiv a(a+1)\cdots(a+m-1)$. This form of $\pi(m)$
allows one to easily calculate moments of the molecule number from the
partition function $Z$ as in equilibrium statistical mechanics, e.g.
$s=\av{m}=c_0 dZ/dc_0 $, and thence the outgoing flux, $c=c_0-\beta
s$. Using the fact that $Z$ can be written explicitly in terms of
hypergeometric functions, we find that the noise index
grows with $\beta$ as $\eta_s^2 \simeq v/(K c_0) +
\beta/c_0$. The distribution function is given in
supporting figure~1 for several values of $\beta$.
%


Generalizing the above to a directed pathway, we allow for $\beta$,
as well as for $v_{\max}$ and $K$, to be $i$-dependent. The
decoupling property \eqref{e.product} does not generally hold in the
non-conserving case
\cite{Evans05}. However, in this case  the stationary distribution
still seems to be well approximated by a product of the
single-metabolite functions $\pi_i(m)$ of  the form~\eqref{e.pib},
with $c_0/\beta \to c_{i-1}/\beta_i$. This is supported again by the excellent
agreement between noise indices  obtained by
numerical simulations and analytic calculations using the
product measure Ansatz, for linear pathways with dilution of intermediates;
see figure~\ref{f.bars}(c). In this case, change in the level of the first enzyme
does "propagate" to the downstream nodes. But this is not a ``noise propagation''
effect, as the mean fluxes $\av{c_i}$ at the different nodes are already
 affected. (To illustrate the effect of leakage, the simulation used parameters that
 corresponded to a huge leakage current which is $20\%$ of the flux. This is
 substantially larger than typical leakage encountered, say due to growth-mediated
 dilution, and we do not expect propagation effects due to leakage to be significant
 in practice.)


\section{Interacting pathways}

The metabolic network in a cell
is composed of pathways of different topologies. While linear
pathways are abundant, one can also find circular pathways (such as
the TCA cycle), converging pathways and diverging ones. Many of
these can be thought of as a composition of interacting linear
pathways. Another layer of interaction is imposed on the system
due to the allosteric regulation of enzyme activity by
intermediate metabolites or end products.
To what extent can our results for a linear pathway be
applied to these more complex networks? Below we address this
question for a few of the frequently encountered cases. To simplify
the analysis, we will consider only directed pathways and suppress
the dilution/leakage effect.

\subsection{Cyclic pathways}
We first address the cyclic pathway, in which the metabolite $S_L$
is converted into $S_1$ by the enzyme $E_L$.
Borrowing a  celebrated result
for queueing networks \cite{Jackson57} and mass transfer models
\cite{Spitzer71}, we note that the decoupling property \eqref{e.product}
described above for the linear directed pathway also holds exactly
even for the cyclic pathways\footnote{In fact, the decoupling property holds
for a general network of directed single-substrate reactions,
even if the network contains cycles.}. This result is surprising
mainly because the Poissonian nature of the ``incoming'' flux
assumed in the analysis so far is lost, replaced in this case by
a complex expression, e.g.,  $w^{(L)}_{m_L}\cdot \pi_L(m_L)$.

 In an
isolated cycle the total concentration of the metabolites, $s_{\rm
tot}$ -- and not the flux  -- is predetermined. In this case, the
flux $c$ is give by the solution to the equation
\begin{equation}
s_{\rm tot} = \sum_{i=1}^{L}s_i(c) =
\sum_{i=1}^{L}\frac{cK_i}{v_i-c}\;.
\end{equation}
Note that this equation can always be satisfied by some positive $c$ that is smaller than all $v_i$'s.
In a cycle that is coupled to other branches of the network, flux
may be governed by metabolites going into the cycle or taken from
it. In this case, flux balance analysis will enable determination of
the variables $z_i$ which specify the probability distribution
\eqref{e.pi}.


\subsection{End-product inhibition}
 Many biosynthesis pathways
couple between supply and demand by a negative feedback
\cite{Coli,Michal}, where the end-product inhibits the first
reaction in the pathway or the transport of its precursor;
see, e.g., the dashed lines in figure~\ref{f.pathway}.  In this way, flux is
reduced when the end-product builds up. In branched
pathways this may be done by regulating an enzyme immediately
downstream from the branch-point, directing some of the flux towards
another pathway.

To study the effect of end-product inhibition, we consider
inhibition of the inflow into the pathway. Specifically, we model the
probability at which substrate molecules arrive at the pathway by a
stochastic process with exponentially-distributed waiting time, characterized by the rate
$\alpha(m_L) = {c_0}\left[{1+ (m_L/K_I)^{h}}\right]^{-1}$,
where $c_0$ is the maximal influx (determined by availability of the
substrate either in the medium or in the cytoplasm), $m_L$ is the number
of  molecules of the end-product ($S_L$),  $K_I$ is the
dissociation constant of the interaction between the first enzyme $E_0$
and $S_L$,  and $h$ is a Hill coefficient describing the cooperativity
of interaction between $E_0$ and $S_L$.
Because $m_L$ is a stochastic variable itself, the incoming flux is described
by a nontrivial stochastic process which is manifestly non-Poissonian.

The steady-state flux is now
\begin{equation}
c = \av{\alpha(m_L)} =
c_0 \cdot \left\langle\left[1+ (m_L/K_I)^{h}\right]^{-1}\right\rangle\;.
\label{e.scc}
\end{equation}
This is an implicit equation for the flux $c$, which also appears in
the right-hand side of the equation through the distribution $\pi(m_1, ...,m_L)$.

By drawing an analogy between feedback-regulated pathway and a cyclic pathway,
we conjecture that metabolites in the former should be effectively uncorrelated.
The quality of this approximation is expected to become better in cases where
the ration between the influx rate $\alpha(m_L)$ and the outflux rate $w_{m_L}$ is typically
$m_L$ idependent. Under this assumption, we approximate
the distribution function by the product measure \eqref{e.product},
with the form of the single node distributions given by \eqref{e.pi}.
Note that the conserved flux then depends on the properties of the enzyme processing the
last reaction, and in general should be influenced by the fluctuations in the controlling metabolite.
In this sense, these fluctuations propagate throughout the pathway at the level of the mean flux.
This should be expected from any node characterized by a high control coefficient \cite{Fell97}.

Using this approximate form, Eq.~\eqref{e.scc} can be solved
self-consistently to yield $c(c_0)$, as is shown explicitly in Supp. Mat. for $h=1$.
The solution obtained is found to be in excellent agreement with numerical simulation
(Supporting figure~2a). The quality of the product measure approximation is further
scrutinized by
comparing the noise index of each node upon increasing
the enzyme level of the first node 5-fold. Figure~\ref{f.bars}(c) shows clearly that
the effect of changing enzyme level does not propagate to other nodes.
While being able to accurately predict the flux and mean metabolite level at each node,
the predictions based on the product measure are found to be under-estimating
the noise index by up to 10\% (compare bars and symbols). We conclude that in this case
correlations between metabolites do exist, but not dominate. Thus analytic expressions dervied
from the decorrelation assumption can be useful even in this case (see supporting figure~2b).


\subsection{Diverging pathways} Many metabolites serve as substrates for several
different pathways. In such cases, different enzymes can bind to the substrate, each
catabolizes a first raction in a different pathway. Within our scheme, this can be modeled
by allowing for a metabolite $S_i$ to be converted to metabolite $S^\mathtt{I}_1$ with rate $w^\mathtt{I}_{m_i}=v^\mathtt{I}m_i/(m_i+K^\mathtt{I}-1)$ or to metabolite $S^\mathtt{II}_1$ with rate $w^\mathtt{I}_{m_i}=v^\mathtt{II}m_i/(m_i+K^\mathtt{II}-1)$. The paramters $v^{\mathtt{I},\mathtt{II}}$ and $K^{\mathtt{I},\mathtt{II}}$ characterize the two different enzymes.

Similar to the case of reversible reactions, the steady-state distribution is given exactly
by a product measure only if $w^\mathtt{I}_{m_i}/w^\mathtt{II}_{m_i}$ is a constant, independent of $m_i$ (namely when $K^\mathtt{I}=K^\mathtt{II}$). Otherwise, we expect it to hold in a range of alternative scenarios, as described for reversible pathways.

Considering a directed pathway with a single branch point, the distribution
\eqref{e.pi} describes exactly all nodes upstream of that point. At the branchpoint, one
replaces $w_m$ by $w_m=w^\mathtt{I}_m+w^\mathtt{II}_m$, to obtain the distribution function
\begin{equation}
\pi(m)=\frac{c^m}{Z}\frac{(K^\mathtt{I})_m(K^\mathtt{II})_m}{m!((K^\mathtt{I}v^\mathtt{II}+K^\mathtt{II}v^\mathtt{II})/(v^\mathtt{I}+v^\mathtt{II}))_m}\;.
\label{e.bpt}\end{equation}
From this distribution one can obtain the fluxes going down each one of the two
branching pathway, $c^{\mathtt{I},\mathtt{II}}=\sum w^{\mathtt{I},\mathtt{II}}_m\pi(m)$.
Both fluxes depend on the properties of {\em both} enzymes, as can be seen from \eqref{e.bpt}, and
thus at the branch-point the two pathways influence each other \cite{LaPorte84}.
Moreover, fluctuations at the branch point to propagate into the branching pathways already
at the level of the mean flux. This is consistent with the fact that the branch node is expected to be
characterized by a high control coefficient \cite{Fell97}.

While different metabolite upstream and including the branch point are uncorrelated, this is not
exactly true for metabolites of the two branches.
Nevertheless, since these pathways are still directed,
we further conjecture that metabolites in the two branching
pathways can still be described, independently, by the probability distribution \eqref{e.pi},
with $c$ given by the flux in the relevant branch, as calculated from \eqref{e.bpt}.
Indeed, the numerical results of figure~\ref{f.bars}(e)
strongly support this conjecture. We find
that changing the noise properties of a metabolite in the upstream pathway do not propagte to those of the branching pathways.

\subsection{Converging pathways -- combined fluxes} 
\begin{figure}[t]
\centerline{\includegraphics{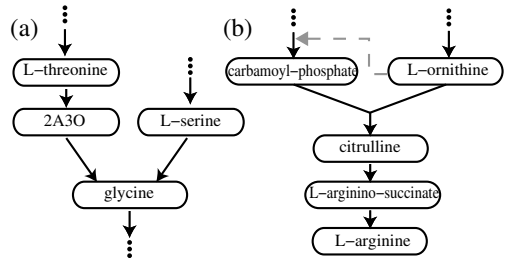}}
\caption{Converging pathways. (a) Glycine is synthesized in two
independent pathways. (b) Citrulline is synthesized from products
of two pathways. Abbreviations: 2A3O, 2-Amino-3-oxobutanoate.} \label{f.conv}
\end{figure}

We next examine
the case where two independent pathways result in synthesis of
the same product, $P$. For example, the amino acid glycine is the
product of two (very short) pathways, one using threonine and the
other serine as precursors (figure~\ref{f.conv}(a)) \cite{Coli}.
With
only directed reactions, the different metabolites in the combined
pathway -- namely, the two pathways producing $P$ and a pathway
catabolizing $P$ -- remain decoupled. The simplest way to see this
is to note that the process describing the synthesis of $P$, being
the sum of two Poisson processes, is still a Poisson process. The
pathway which catabolizes $P$ is therefore statistically identical
to an isolated pathway, with an incoming flux that is the sum of the
fluxes of the two upstream pathways. More generally, the Poissonian
nature of this process allows for different pathways to dump or take
from common metabolite pools, without generating complex correlations
among them.

\subsection{Converging pathways -- reaction with two
fluctuating substrates}
As mentioned above, some reactions in a biosynthesis pathway involve
side-reactants, which are assumed to be abundant (and hence at a
constant level). Let us now discuss briefly a case where this
approach fails. Suppose that the two products of two linear pathways
serve as precursors for one reaction. This, for example, is the case
in the arginine biosynthesis pathway, where L-ornithine is combined
with carbamoyl-phosphate by ornithine-carbamoyltransferase to create
citrulline (figure~\ref{f.conv}(b))
\cite{Coli}. Within a flux balance model, the net fluxes of both
substrates must be equal to achieve steady state, in which case the
macroscopic Michaelis-Menten flux takes the form
$$c = v_{\max}\frac{[S_1][S_2]}{(K_{M1}+[S_1])(K_{M2}+[S_2])}\;.$$
Here $[S_{1,2}]$ are the steady-state concentrations of the two
substrates, and $K_{M1,2}$ the corresponding MM-constants. However,
flux balance provides only one constraint to a system with two
degrees of freedom.

\begin{figure}[t]
\centerline{\includegraphics{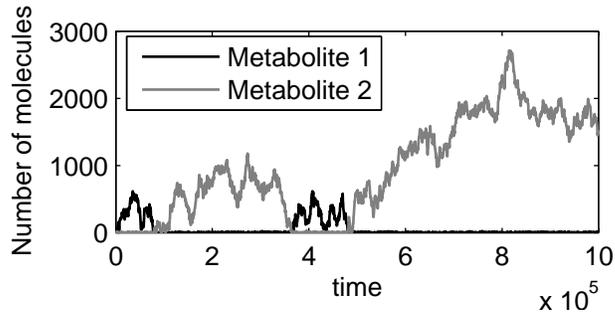}}
\caption{Time course of a two-substrate enzymatic reaction, as
obtained by a Gillespie simulation \cite{Gillespie}. Here $c=3
t^{-1}$, $k_+=5 t^{-1}$ and $k_-=2 t^{-1}$ for both substrates,
$t$ being an arbitrary time unit.} \label{f.twosubstrates}
\end{figure}

In fact, this reaction exhibits no steady state. To see why,
consider a typical time evolution of the two substrate pools
(figure~\ref{f.twosubstrates}). Suppose that at a certain time one
of the two substrates, say $S_1$, is of high molecule-number
compared with its equilibrium constant, $m_1 \gg K_1$. In this case,
the product synthesis rate is unaffected by the precise value of
$m_1$, and is given approximately by $v_{\max} m_2/ (m_2+K_2)$.
Thus, the number $m_2$ of $S_2$ molecules can be described by the
single-substrate reaction analyzed above, while $m_1$ performs a
random walk (under the influence of a weak logarithmic potential),
which is bound to return, after some time $\tau$, to values
comparable with $K_1$. Then, after a short transient, one of the two
substrates will become unlimiting again, and the system will be back
in the scenario described above, perhaps with the two substrates
changing roles (depending on the ratio between $K_1$ and $K_2$).

Importantly, the probability for the time $\tau$ during which one of
the substrates is at saturating concentration scales as
$\tau^{-3/2}$ for large $\tau$. During this time the substrate pool
may increase to the order $\sqrt{\tau}$.  The fact that $\tau$ has
no finite mean implies that this reaction has no steady state. Since
accumulation of any substrate is most likely toxic, the cell must
provide some other mechanism to limit these fluctuations. This may
be one interpretation for the fact that within the arginine
biosynthesis pathway, L-ornithine is an enhancer of
carbamoyl-phosphate synthesis (dashed line in figure~\ref{f.conv}(b)).

In contrast, a steady-state always exists if the two metabolites experience linear degradation,
as this process prevents indefinite accumulation. However, in general one expects enzymatic reactions
to dominate over futile degradation. In this case,  equal in-fluxes of the two substrates result in large fluctuations, similar to the ones described above \cite{Elf03}.


\section{Discussion}
In this work we have characterized stochastic fluctuations  of
metabolites for dominant simple motifs of the metabolic network
in the steady state.
Motivated by the analogy between the directed biochemical pathway
and the mass transfer model
or, equivalently, as the queueing network, we show that the
intermediate metabolites in a linear pawthway -- the key motif of the biochemical netrowk --
are statistically independent.
We then extend this result to a wide
range of pathway structures.
Some of the results (e.g., the directed linear, diverging and cyclic pathways)
have been proven previously in other contexts.
In other cases (e.g., for reversible reaction, diverging pathway
or with leakage/dilution),
the product measure is not exact.
Nevertheless, based on insights from the exactly solvable models, we
conjecture that it still describes faithfuly the statistics of the pathway.
Using the product measure as an {\it Ansatz}, we obtained
quantitative predictions which turned out to be in excellent agreement
with the numerics (figure~\ref{f.bars}). These results suggest that the product measure may
be an effective starting point for quantitative, non-perturbative
analysis of (the stochastic properties) of these circuit/networks. We hope this study will stimulate
further analytical studies of  the large variety of pathway topologies in metabolic networks, as well as in-depth mathematical analysis of the conjectured results.
Moreover, it will be interesting to explore the applicability of the present
approach to other cellular networks, in particular,
stochasticity in protein signaling networks \cite{Brent05},
whose basic mathematical structure is also a set of interlinked Michaelis-Menton
reactions.

Our main conclusion, that the steady-state fluctuations in each
metabolite depends only on the properties of the reactions consuming
that metabolite and not on fluctuations in other upstream
metabolites, is qualitatively different from conclusions obtained
for gene networks in recent studies, e.g., the ``noise addition
rule'' \cite{Paulsson04, Shibata05} derived from the {\it
Independent Noise Approximation}, and its extension to cases where
the singnals and the processing units interact \cite{Wolde06}. The
detailed analysis of \cite{Wolde06}, based on the {\it Linear Noise
Approximation} found certain anti-correlation effects which reduced
the extent of noise propagation from those expected by ``noise
addition'' alone \cite{Paulsson04,Shibata05}. While the specific
biological systems studied in \cite{Wolde06} were taken from protein
signaling systems, rather than metabolic networks, a number of
systems studied there are identical in mathematical structure to
those considered in this work.  It is reassuring to find that
reduction of noise propagation becomes complete (i.e., no noise
propagation) according to the analysis of \cite{Wolde06}, also, for
Poissonian input noise where direct comparisons can be made to our
work (ten Wolde, private communication). The cases in which residue
noise propagation remained in \cite{Wolde06}, corresponded to
certain ``bursty'' noises which is non-Poissonian. While bursty
noise is not expected for metabolic and signaling reactions, it is
nevertheless important to address the extent to which the main
finding of this work is robust to the nature of stochasticity in the
input and the individual reactions. The exact result on the cyclic
pathways and the numerical result on the directed pathway with
feedback inhibition suggest that our main conclusion on statistical
independence of the different nodes extends significantly beyond
strict Poisson processes. Indeed, generalization that preserve this
property include classes of transport rules and extended topologies
\cite{Evans04,Greenblatt06}.

The absence of noise propagation for a large part of the metabolic network
allows intermediate metabolites
to be shared freely by multiple reactions in multiple pathways, without the need
of installing elaborate control mechanisms.
In these systems, dynamic fluctuations (e.g., stochasticity in enzyme expression
which occurs at a much longer time scale) stay local to the node,
and are shielded from triggering system-level
failures (e.g., grid-locks). Conversely, this property allows convenient implementation
of controls on specific node of pathways, e.g., to limit the pool of a specific toxic
intermediate, without the concern of elevating fluctuations in other nodes.
We expect this to make the evolution of metabolic network less
constrained, so that the system can modify its local properties nearly freely
in order to adapt to environmental or cellular changes. The optimized pathways
can then be meshed smoothly into the overall metabolic
network, except for junctions between pathways where complex
fluctuations not constrained by flux conservation.

In recent years, metabolomics, i.e., global metabolite
profiling, has been suggested as a tool to decipher
the structure of the metabolic network \cite{Arkin95,Weckwerth02}.
Our results suggest that in many cases, steady-state fluctuations
do not bare information about the pathway structure. Rather,
correlations  between metabolite fluctuations may be, for example,
the result of fluctuation of a common enzyme or coenzyme, or reflect
dynamical
disorder \cite{English05}. Indeed, a bioinformatic study found no
straightforward connection between observed correlation and the
underlying reaction network \cite{Steuer03}. Instead, the
response to external perturbation  \cite{Arkin97,Arkin95,Vance02}
may be much more effective in shedding light on
the underlying structure of the network, and may be used to
study the morphing of the network under different conditions.
It is important to note that all results described here are applicable only
to systems in the steady state; transient responses such as
the establishment of the steady state and the response
to external perturbations will likely exhibit complex temporal
as well as spatial correlations. Nevertheless, it is possible
 that some aspects of the response function may be
 attainable from the steady-state fluctuations through non-trivial
fluctuation-dissipation relations as was shown for other
related nonequilibrium systems \cite{Liggett,Forster77}.


\ack
We are grateful to Peter Lenz and Pieter Rein ten Wolde for discussions.
This work was
supported by NSF through the PFC-sponsored Center for Theoretical
Biological Physics (Grants No. PHY-0216576 and PHY-0225630).  TH
acknowledges  additional support by NSF Grant No. DMR-0211308.

\appendix
\section*{Supporting Material}

\section{Microscopic model}  Under the assumption of fast
equlibration between the substrate and the enzyme, the probability of
having $N_{SE}$ complexes given $m$ substrate molecules and $N_E$
enzymes is given by equation \meqref{3} of the main text. To write the
partition function explicitly, we define $u(x)=  U(x,1-m-N_E;-K)$,
where $U$ denotes the Confluent Hypergeometric function \cite{Abramowitz72}. One can then
write the partition sum as $Z_{m,N_E}=(-K)^{-N_E} u(-m)$.  The
turnover rate is then given by $w_m = \frac{k_2 N_E}{V} [-m\, u(1-m)]/
[u(-m)]$, which can be approximated by Equation \meqref{4}.

\section{Influx of metabolites}
A metabolic reaction {\it in vivo} can be described as turnover of an
incoming flux of substrate molecules, characterized by a Possion
process with rate $c$, into an outgoing flux. To find the probability
of having $m$ substrate molecules we write down the Master equation,
\begin{equation}
\frac{d}{dt}\pi(m) = \left[c(a-1)+(\hat{a}-1)w_m\right]\pi(m)
=c[\pi(m-1)-\pi(m)]+[w_{m+1}\pi(m+1)-w_m\pi(m)]
\;,\label{e.influx}
\end{equation}
where we took the opportunity to define the lowering and raising
operators $a$ and $\hat{a}$, which -- for any function $h(n)$ --
satisfy $a h(n)=h(n-1)$, $a h(0)=0$, and $\hat{a} h(n) = h(n+1)$.
The first term in this equation is the influx, and the second is the
biochemical reaction. The solution of this steady state equation is
of the form $\pi(m) \sim c^m / \prod_{k=1}^{m}{w_k}$ (up to a
normalization constant), as  can be verified by plugging it into the
equation, 
\begin{equation}
\left[c\left(\frac{\pi(m-1)}{\pi(m)}-1\right)+\left(\frac{\pi(m+1)}
{\pi(m)}w_{m+1}-w_m\right)\right] = c \left(\frac{w_m}{c}-1\right)+
\left(\frac{c}{w_{m+1}}w_{m+1}-w_m\right)=0.
\end{equation}
Using the approximate form of $w_m$, as given in \meqref{4}, the
probability  $\pi(m)$ takes the form,
\begin{equation}
\pi(m) = {m+K+(N_E-1)\choose m}(1-z)^{K+N_E}z^m\;,
\label{e.Spi}
\end{equation}
as given in equation~\meqref{5} of the main text.

\section{Directed linear pathway}
We now derive our key results, equation~\meqref{8} 
(The result has been derived  previously in the context of
queueing networks  \cite{Taylor98},
and of mass-transport systems \cite{Levine05}). To this end we
write the Master equation for the joint probability function $\pi
\equiv \pi(m_1,m_2,\cdots,m_L)$,
\begin{equation}
\frac{d}{dt}{\pi} = \left[c (a_1-1) + \sum_{i=1}^{L-1} (\hat{a}_i a_{i
+1}-1) w^{(i)}_{m_i}+(\hat{a}_{L}-1)w^{(L)}_{m_L}\right]\pi\;,
\label{e.stm}
\end{equation}
which generalizes \eqref{e.influx}.
As above, $a_i$ and $\hat{a_i}$ are lowering and raising operators,
acting on the number of $S_i$ molecules. The first term in this
equation is the incoming flux $c$ of the substrate, and the last term
is the flux of end product. Let us try to solve the steady-state
equation by plugging a solution of the form $\pi(m_1,m_2,\cdots,m_L)=
\prod g_i(m_i)$, yielding
\begin{equation}
\\[-24pt]c[\frac{g_i(m_1-1)}{g_1(m_1)}-1]+\sum_{i=1}^{L-1}[w^{(i)}_
{m_i+1}\frac{g_i(m_i+1)g_{i+1}(m_{i+1}-1)}{g_i(m_i)g_{i+1}(m_{i+1})}-
w^{(i)}_{m_i}]
+[w^{(L)}_{m_L+1}\frac{g_L(m_L+1)}{g_L(m_L)}-w^{(L)}_{m_L}]=0\;.
\end{equation}
Motivated by the solution to \eqref{e.influx}, we try to satisfy this
equation by choosing $g_i(m) = c^m /\prod_{k=1}^{m}w^{(i)}_k$. With
this choice we have $g(m+1)/g(m)=c/w_{m+1}$ and $g(m-1)/g(m)=w_m/c$.
It is now straightforward to verify that indeed
\begin{equation}
c\left(\frac{w^{(1)}_{m_1}}{c}-1\right)+\sum_{i=1}^{L-1}\left(w^{(i
+1)}_{m_{i+1}}-w^{(i)}_{m_i}\right)+\left(c-w^{(L)}_{m_L}\right)=0\;.
\end{equation}
Finally, in our choice of $g_i(m)$ we replace $w^{(i)}_m$ by the MM-
rate $v_i m_i / (m_i+K_i)$, and find that in fact $g_i(m) = \pi_i(m)
$, namely
\begin{equation}
\pi(m_1,m_2,\ldots,m_L)= \prod_{i=1}^{L}\pi_i(m_i)\;,
\label{e.Sproduct}
\end{equation}
as stated in \meqref{8}.

\section{End-product inhibition}
Equation~\meqref{13} of the main text is a self-consistent equation for the steady-
state flux $c$ through a pathway regulated via end-product inhibition. 
Using considerations analogous to what led to the
exact result on the product measure distribution for the cyclic pathways,
we conjecture that even for the present case of end-product
inhibition, the distribution function can still be approximated by the product measure
\eqref{e.Sproduct} with the form of the single node distributions given by \eqref{e.Spi}.
The flux $c$ enters the calculation of the average on
the right-hand side through the probability function $\pi(m)$. Solving this equation for $c$ yields the steady state current,
and consequently determines the mean occupancy and standard
deviation of all intermediates.

To verify the validity of this conjecture, and to demonstrate its application, we consider the case $h=1$. In this case one can carry the sum, and
find
\begin{eqnarray}
\label{e.feedback}
  c&=& \sum_{m_L=0}^{\infty} c_0 \left[ 1+ (m_L/K_I)^{h}\right]^{-1}
\pi_L(m_L)\\
&=& c_0 (1-z)^{K_L} {_2F_1}(K_I, K_L ; K_I+1 ; z) \nonumber
\end{eqnarray}
with $z=c/v_L$ and $_2F_1$ the hypergeometric function \cite{Abramowitz72}.
This equation was solved numerically, and 
plotted in  supporting figure~\ref{f.SuppFeedback}(a) for some values of $K_I$ and $K_L$. 
Note that  predictions based on the product measure  (lines)
are in excellent agreement with the results of numerical simulation (circles)
for the different sets of parameters tried.

Results obtained from equation \eqref{e.feedback} can be used, for example, to compare 
the flux that flows through the noisy pathway with
the mean-field flux $c_{\rm MF}$,
obtained when one ignores fluctuations in $m_L$, i.e.,
\begin{equation}
c_{\rm MF} = \frac{c_0}{1+(s_L/K_I)^h}\;. \label{e.mf}
\end{equation}
The fractional difference $\delta c=(c-c_{\rm MF})/c_{\rm MF}$ is plotted in 
supporting figure~\ref{f.SuppFeedback}(b). 
The results show that
number fluctuations in the end-product always {\em increase} the  flux in the
pathway since $\delta c > 0 $ always. Quantitatively, this increase
can easily be several percent.
For large $c_0$, a simplifying expression can be derived by using an asymptotic expansion
of the hypergeometric function \cite{Abramowitz72}. For example, when $K_I < K_L$,
\begin{equation}
(1-z)^{K_L} {_2F_1}(K_I, K_L ; K_I+1 ; z) \sim \frac{v_L K_L}{1+K_L-K_I} 
\;,
\end{equation}
which yields 
\begin{equation}
\frac{c-c_{\rm MF}}{c_{\rm MF}} \sim \frac{1}{K_I}\frac{v_L}{c_0}\;.
\label{e.asympt}
\end{equation}
Thus the effect of end-product fluctuations on the current is
enhanced by stronger binding of the inhibitor (smaller $K_I$),
 as one would expect. We note that obtaining these predictions from Monte-Carlo simulation is 
 rather difficult, given the fact that one is interested here in sub-leading quantities.

 \begin{figure}[t]
\centerline{\includegraphics{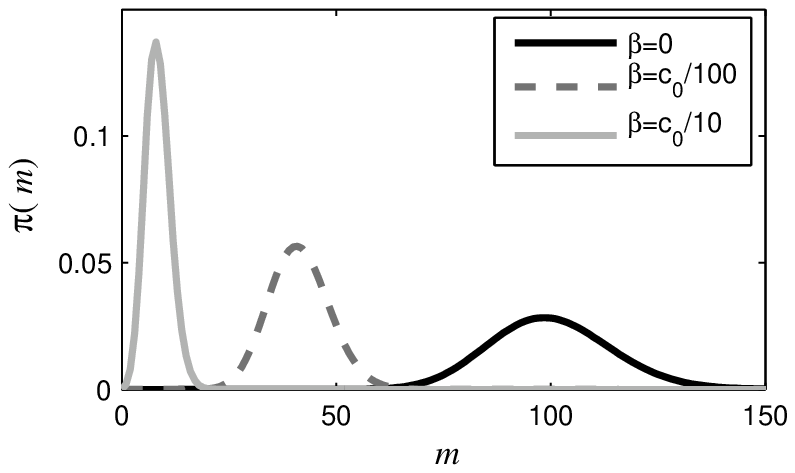}}
\caption{The steady-state distribution $\pi(m)$ of a metabolite, that experiences enzymatic reaction (with rate $w_m=v m/(m+K-1)$) and linear degradation (with rate $\beta m$), as given by equation~\meqref{10} of the main text. Here $K=100$ and $v = 2 c_0$.}
\label{f.SuppDilution}
\end{figure}

\begin{figure}[t]
\begin{flushleft}
{\includegraphics{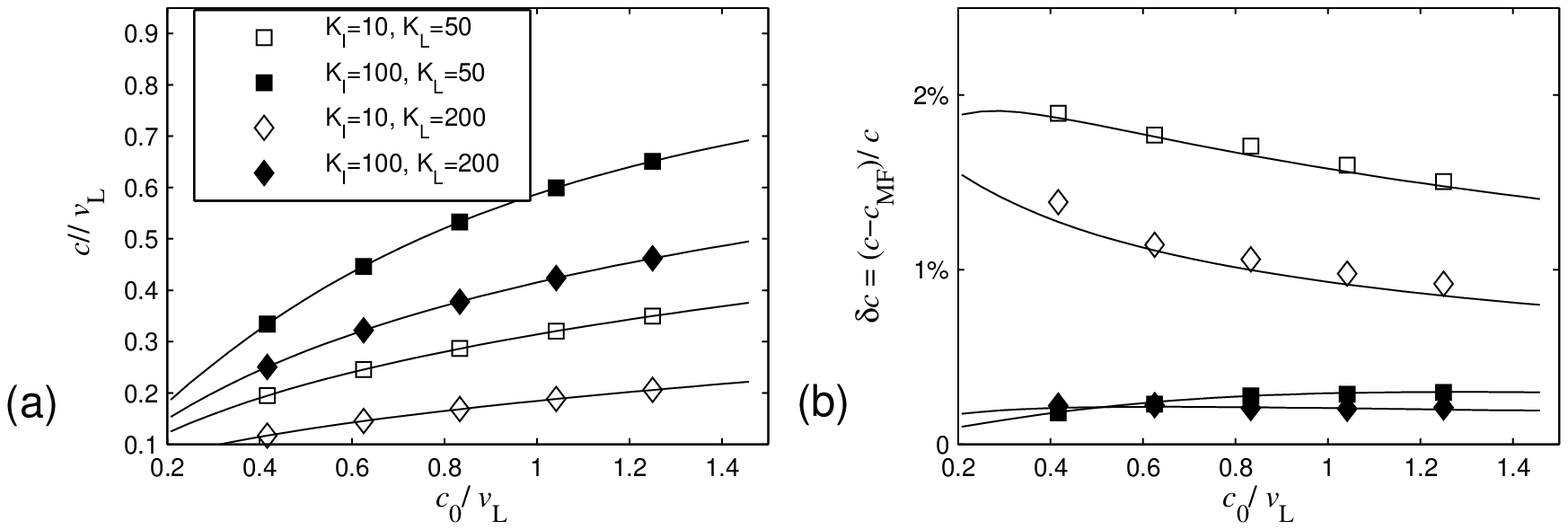}}
\end{flushleft}
\caption{Pathway with end-product inhibition. The influx rate is taken to be $c_0/(1+m_L/K_I)$, and thus the steady-state flux is given by equation~\meqref{12} of the main text, with $h=1$. (a) Assuming that different metabolites in the pathway remain decoupled even in the presence of feedback regulation, \meqref{12} can be approximated by \eqref{e.feedback}. Numerical solutions of equation~\eqref{e.feedback} (lines) are compared with Monte-Carlo simulations (symbols).  Values of parameters are chosen randomly such that $100<K_i<1000$ and $c<v_i<10 c$. For the data presented here, $v_L=2.4 c$. 
We find that \eqref{e.feedback} yields excellent prediction for the steady-state flux.
(b) Neglecting fluctuations altogether yields a mean-field approximation for the flux, $c_{\rm MF}$, given in \eqref{e.mf}. For the same data of (a), we plot  the fractional difference $\delta_c = (c-c_{\rm MF})/c$. 
We find that steady-state flux is increased by fluctuations, and thus taking fluctuations into account (even in an approximate manner) better predicts the steady-state flux. }
\label{f.SuppFeedback}
\end{figure}

\cleardoublepage

\end{document}